\documentclass[aps,pra,twocolumn,amssymb,amsfonts,floatfix,showpacs,]{revtex4}
\usepackage{bm}
\usepackage{dcolumn}% Align table columns on decimal point
\usepackage{bm}% bold math
\usepackage{mathbbol}
\usepackage{graphicx}
\usepackage{epstopdf} 
\usepackage{amsmath,amsthm}
\usepackage{times}
\usepackage{color}
\usepackage{lipsum}

\begin{document}
\title{Structure of eigenstates and quench dynamics at an excited state quantum phase transition}

\author{Lea F. Santos}
\affiliation{Department of Physics, Yeshiva University, New York, New York 10016, USA \\
ITAMP, Harvard-Smithsonian Center for Astrophysics, Cambridge, MA 02138, USA}
\author{Francisco P\'erez-Bernal}
\affiliation{Departamento de F\'{\i}sica Aplicada, Facultad de Ciencias Experimentales, Universidad de Huelva, Campus del Carmen, 
Avda.\ de las Fuerzas Armadas s/n, 21071 Huelva, SPAIN}

\date{\today}

\begin{abstract}
We study the structure of the eigenstates and the dynamics of a system that undergoes an excited state quantum phase transition (ESQPT).  The analysis is performed for two-level pairing models characterized by a $U(n+1)$ algebraic structure. They exhibit a second order phase transition between two limiting dynamical symmetries represented by the $U(n)$ and $SO(n+1)$ subalgebras. They are, or can be mapped onto, models of interacting bosons.  We show that the eigenstates with energies very close to the ESQPT critical point, $E_{\text{ESQPT}}$, are highly localized in the $U(n)$-basis.  Consequently, the dynamics of a system initially prepared in a $U(n)$-basis vector with energy ${\cal{E}} \sim E_{\text{ESQPT}}$ may be extremely slow. Signatures of an ESQPT can therefore be found in the structures of the eigenstates and in the speed of the system evolution after a sudden quench. Our findings can be tested experimentally with trapped ions.
\end{abstract}

% 05.30.Rt 	Quantum phase transitions (see also 64.70.Tg Quantum phase transitions in specific phase transitions; and 73.43.Nq Quantum phase transitions in Quantum Hall effects) 
% 21.60.Fw 	Models based on group theory
% 64.70.qj 	Dynamics and criticality
% 64.70.Tg 	Quantum phase transitions (for quantum Hall effects aspects, see 73.43.Nq in electronic structure of surfaces, interfaces, thin films, and low dimensional structures)

\maketitle

%%%%%%%%%%%%%%%% INTRODUCTION %%%%%%%%%%%%%%%%%%%%%%%%%

{\em Introduction.--} Quantum phase transitions (QPTs) occur at zero temperature. They correspond to an abrupt change in the character of the ground state of a system when a control parameter passes a critical point~\cite{SachdevBook}. The subject, which permeates condensed matter and nuclear physics, has become one of the highlights of experiments with cold gases, where transitions from a superfluid to a Mott insulator~\cite{Greiner2002} and from a normal to a superradiant phase~\cite{Baumann2010} have been observed. The investigations are not restricted to the properties of the ground state, but extend also to the dynamics of systems undergoing QPTs. In this context, one finds studies about the quantum analogue of the Kibble-Zurek mechanism~\cite{Zurek2005}, as well as
the relaxation time~\cite{Barmettler2009,Heyl2013,Heyl2014,Canovi2014}, revivals~\cite{romera2013,romera2015}, and temporal fluctuations~\cite{Venuti2010b,Venuti2014} at critical points.

Recently, the concept of ground state QPT has been generalized to encompass also QPTs occurring at excited states. These so-called ESQPT refer to a singularity in the energy spectrum caused by the clustering of excited levels at a critical energy~\cite{Cejnar2006,cejnarPRL,Caprio2008,Cejnar2008}. This critical point can be reached either for a constant excited energy by varying the control parameter(s) or by fixing the latter and increasing the energy.  ESQPTs have been investigated for a broad class of many-body quantum systems, such as the Lipkin-Meshkov-Glick (LMG)~\cite{Cejnar2009,Fernandez2009,Yuan2012}, the molecular vibron~\cite{Caprio2008,Bernal2008}, the nuclear interacting
boson~\cite{Cejnar2009}, the Jaynes-Cummings~\cite{Fernandez2011,Fernandez2011b}, the Dicke~\cite{Fernandez2011,Fernandez2011b,Brandes2013}, and the kicked-top \cite{kicked_top} models.   Experimentally, signatures of ESQPTs were found in molecular systems~\cite{Larese1, Larese2, jcp_bending, Winnewisser2005, zobov2006}, superconducting microwave billiards~\cite{Dietz2013}, and spinor condensates~\cite{Zhao2014}.

In terms of dynamics, it has been shown that an ESQPT leads to random oscillations of the survival probability in isolated systems~\cite{Fernandez2011}, to singularities in the evolution of observables~\cite{Engelhardt2015}, and to maximal decoherence in open systems~\cite{Relano2008,Fernandez2009}. Despite these works, studies of the effects of ESQPTs on systems' evolutions are still scarce. 

In this Rapid Communication, we provide insights into the dynamics of an isolated many-body quantum system that undergoes an ESQPT. The system is prepared in an eigenstate of an initial Hamiltonian $\hat {\cal H}_I$. The evolution starts after the sudden quench of a control parameter $\xi$ that changes $\hat {\cal H}_I$ into a new final Hamiltonian $\hat {\cal H}_F$ described by a $U(n+1)$ algebraic structure,
\begin{equation}
\hat {\cal H}_F = (1- \xi) \hat {\cal H}_{U(n)}  + \dfrac{\xi}{N} \hat {\cal H}_{SO(n+1)}  .
\label{ham}
\end{equation}
This model exhibits a second order ground state QPT at $\xi_c = 0.2$, which occurs between the dynamical symmetries (DSs) represented by the $U(n)$ and the $SO(n+1)$ subalgebras~\cite{Cejnar2007}.  It also displays an ESQPT at an energy $E_{\text{ESQPT}}(\xi)$ for $\xi_{\text{ESQPT}} > \xi_c$ \cite{Caprio2008}. The model represents systems of interacting bosons. It is built upon two types of bosons, a scalar and a non-scalar one. $\hat {\cal H}_{U(n)}$ is  the number operator of the non-scalar boson, while $\hat {\cal H}_{SO(n+1)}$ is a two-body (pairing) operator built from the second order invariant operator of the $SO(n+1)$ subalgebra, which is rescaled by the system size, $N$. Such models have been successfully applied to problems of hadronic~\cite{hadrons}, nuclear~\cite{booknuc}, and molecular physics~\cite{IachelloBook}. For $n=1$, they coincide with  the LMG model~\cite{Lipkin1965} in the bosonic form.

We assume that the initial Hamiltonian corresponds to one of the two limits of $\hat {\cal H}_F$, with $\xi=0$ or $\xi=1$. Thus, the initial state $|\Psi(0)\rangle$ is either a basis-vector $|\phi_{U(n)} \rangle$ associated with the DS $U(n+1)\supset U(n)\supset \dots$, which defines the so-called ``spherical'' or ``symmetrical'' phase, or it is a basis-vector $|\phi_{SO(n+1)} \rangle$ associated with the DS $U(n+1)\supset SO(n+1)\supset \dots$, which corresponds to the ``deformed'' or ``broken symmetry'' phase~\cite{Cejnar2007}. We study the evolution of initial states with different values of the energy
\begin{equation}
{\cal E} = \langle \Psi(0) | \hat {\cal H}_F | \Psi(0) \rangle ,
\label{Eini}
\end{equation}
and show that the rate at which these states change in time can be anticipated from the eigenstates structures.

The initial state $|\Psi(0)\rangle = | \phi_{U(n)} \rangle$ with energy ${\cal E}$ closest to $E_{\text{ESQPT}}$ corresponds to the ground state of the $U(n)$ Hamiltonian. The evolution of this state is extremely slow.   This happens because the main contributions to its dynamics stem from very few eigenstates of $\hat {\cal H}_F$ whose energies are exceedingly close to the separatrix that marks the ESQPT. These eigenstates are strongly localized in the ground state of the $U(n)$ Hamiltonian

Differently from ground state QPTs, the characterization of the different phases in ESQPTs is hindered by the fact that the order parameter does not vanish above or below the critical point. However, degeneracy patterns with respect to angular momentum~\cite{Caprio2008,Bernal2010} and the structures of the eigenstates reveal information about the phases~\cite{Caprio2008}. For instance, the distribution of the $U(n)$-components of the eigenstates below and above the separatrix resemble, respectively, the distributions of the $U(n)$ components of the eigenstates of $\hat {\cal H}_{SO(n+1)}$ (eigenstates of the deformed phase [SO(n + 1) DS] ) and of the eigenstates of $\hat {\cal H}_F \sim  \hat {\cal H}_{U(n)}$ for $\xi \rightarrow 0$ (eigenstates close to the spherical phase [$U(n)$ DS]) (see details in Ref.~\cite{Caprio2008}). The aforementioned excited eigenstates at the separatrix are the ones that mark the change of character from one symmetry to the other, being highly localized in the ground state of the spherical configuration.  In contrast, when written in the $SO(n+1)$-basis, the eigenstates close to the separatrix are delocalized and do not show particular structures. As a result, the evolution of $|\Psi(0)\rangle = |\phi_{SO(n+1)} \rangle$ is not much affected by the ESQPT.

The $U(n)$-basis plays a special role in the study of ESQPTs. As we show below, the point of the transition is clearly revealed from the analysis of the structure of the eigenstates in this basis as well as from the dynamics starting from a $U(n)$-basis vector.

%%%%%%%%%%%%%%%% MODEL %%%%%%%%%%%%%%%%%%%%

{\em Model.--} The findings described in this work were numerically confirmed for one-, two-, and three-dimensional vibron models~\cite{Iachello1981,IachelloBook,Iachello1996,Bernal2005,Bernal2008}, characterized respectively by the $U(2)$, $U(3)$, and $U(4)$ algebraic structures, as well as for the $U(2)$ LMG model~\cite{Lipkin1965}. The results were equivalent, so we chose the 3D case to present the illustrations below.

The $U(4)$ vibron model provides an algebraic framework for the full rovibrational spectrum of diatomic molecules~\cite{IachelloBook}.  It has two rotationally invariant DSs: $U(4) \supset U(3) \supset SO(3)$ and  $U(4)
\supset SO(4) \supset SO(3)$. The $U(3)$ limit describes the vibrational spectrum of non-rigid molecules, while the $SO(4)$ limit corresponds to rigid molecules with vanishing vibrational transitions.  The ground state QPT of this model was studied in \cite{Zhang2008}. 

The $U(4)$ dynamical algebra generators are bilinear products of
creation and annihilation operators of the scalar $s$ and the vector
$p_\mu$ boson operators, with $\mu = 0, \pm 1$. The scaled
Hamiltonian, $\hat{\cal H}_{U(4)} = (1-\xi)\hat n + \xi \hat P/N$, is
built from the number operator, $\hat n = \sqrt{3}[p^\dagger\times
\tilde p]^{(0)}$, which is the first order Casimir operator of the
$U(3)$ subalgebra, and the pairing operator, $\hat P=N(N+1)-\hat W^2$,
that contains the second order Casimir operator of the $SO(4)$
subalgebra: $\hat W^2 =\hat D^2 + \hat L^2$, where $\hat L_\mu =
\sqrt{2}[p^\dagger\times \tilde p]^{(1)}_\mu$ and $\hat D_\mu =
i[p^\dagger \times \tilde s + s^\dagger \times \tilde p]^{(1)}_\mu $
\cite{IachelloBook,FrankBook}. The $U(3)$-basis, $|[N] n L \rangle$,
has quantum numbers $n = 0,1, \ldots , N-1, N$ and $L = 0 \mbox{ or }
1, \ldots, n-2, n$. The $SO(4)$-basis, $|[N] w L\rangle$, has quantum
numbers $w = 0 \mbox{ or } 1, \ldots , N-2, N$ and $L = 0, 1, \ldots,
w-1, w$. The matrix elements of the two Casimir operators in any of
the two bases can be found in \cite{IachelloBook,FrankBook}.

%%%%%%%%%%%%%%%% EIGENVALUES %%%%%%%%%%%%%%%%%%%%
\begin{figure}[htb]
\centering
\includegraphics*[width=3.in]{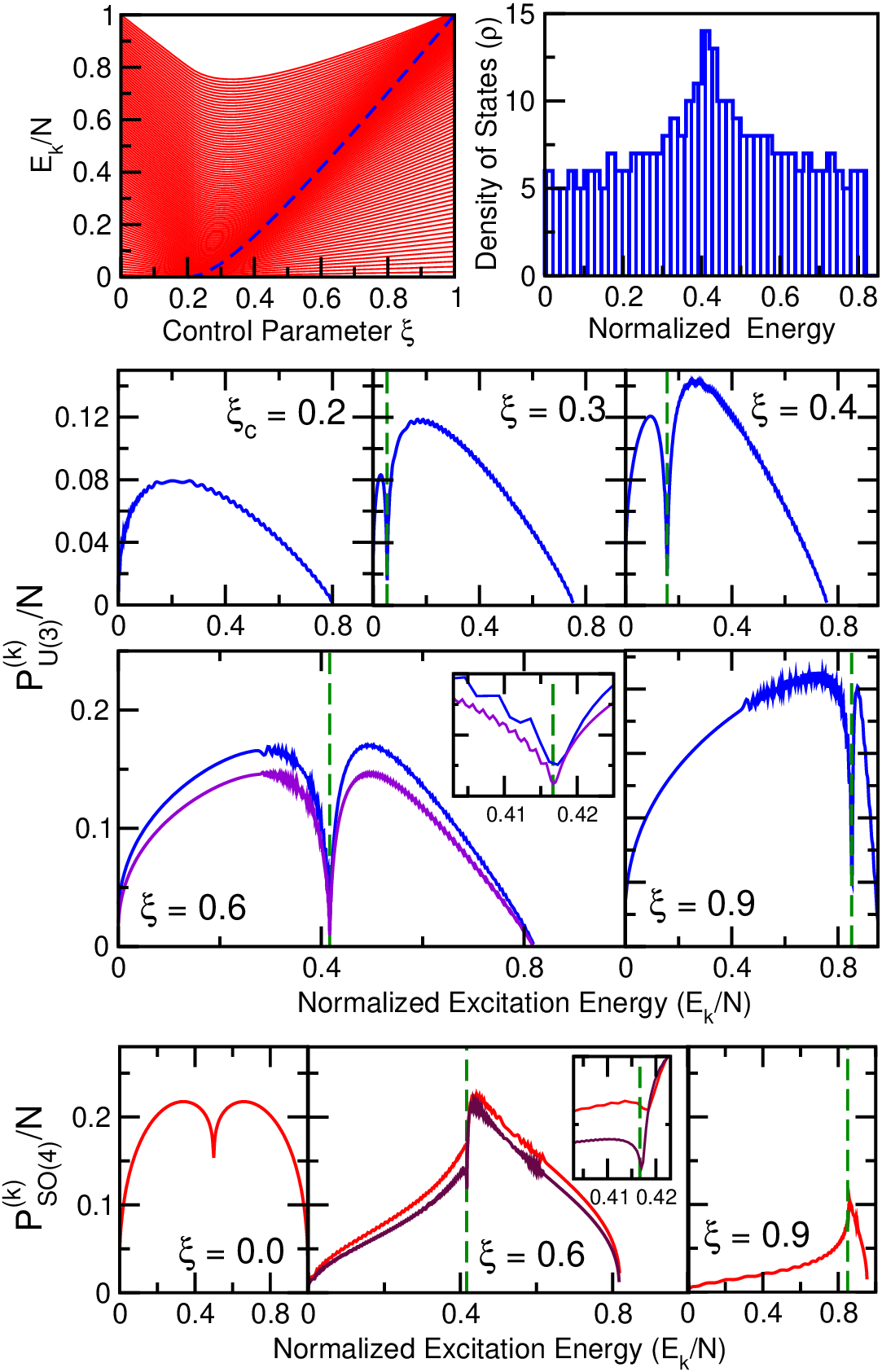}
\caption{(Color online) Top left: normalized excitation energies \textit{vs} $\xi$; $N = 200$. Top right: density of states $\rho$; $N = 400$, $\xi = 0.6$. Middle panels: participation ratio of the eigenstates in the $U(3)$-basis; $N = 600$ [$\xi=0.6$ has also $N=2000$ (bottom curve) and a zoomed in inset]. Bottom panels: same for the $SO(4)$-basis. All panels:  $L = 0$; vertical lines mark $E_{\text{ESQPT}}$ from Eq.~\eqref{eq:separatrix}.}
\label{fig:DOS_IPR}
\end{figure}
{\em Separatrix and density of states.--} For $\xi$ above the critical point, $\xi_c=0.2$,  there appears an energy region in the spectrum of $\hat{\cal H}_{U(4)}$ with a large density of excited levels signali\-zing the ESQPT. This is illustrated in the top left panel of Fig.~\ref{fig:DOS_IPR}, where we plot the normalized excitation energies $E_k/N$ for all levels as a function of the control parameter $\xi$. [Throughout this work, the value of the energy of the ground state is set to zero. In the figures, the units for energy and time are arbitrary.]  The dashed line is the separatrix that divides the states with different physical characters: those closer to the deformed configuration below the separatrix and  those closer to the spherical configuration above it. The equation for
the separatrix as a function of $\xi$ in the  mean field (large $N$) limit is
\begin{equation}
E_{\text{ESQPT}} = (1-5\,\xi_{\text{ESQPT}})^2/(16\,\xi_{\text{ESQPT}}).
\label{eq:separatrix}
\end{equation}
This equation was derived in \cite{Bernal2008,Bernal2010} for the $U(3)$ model and can be extended to other $U(n)$ models~\cite{OurPreparation}.
The large density of states along the separatrix is made evident by the peak at  $E_{\text{ESQPT}}$ in the energy levels histogram in the top right panel of Fig.~\ref{fig:DOS_IPR}.

%%%%%%%%%%%%%%%%%%%%%% EIGENSTATES %%%%%%%%%%%%%%%%%%%%%%%

{\em Structure of the eigenstates.--} To analyze of the structure of the eigenstates of the total Hamiltonian
$\hat{\cal H}_{U(4)}$, we use the participation ratio $P$, which quantifies the level of delocalization of a state in a particularly chosen basis~\cite{notePR}. A large value indicates an extended state in that basis and a small value, a localized state. For an eigenstate written in the $U(3)$-basis, $|\psi_{k}\rangle = \sum_{n=L}^{N} C_n^{(k)} |[N] n
L\rangle_k$,
\begin{equation}
\text{P}^{(k)}_{U(3)} =\frac{1}{\sum_{n} | C_{n}^{(k)}|^4}.
\label{eq:IPR}
\end{equation}
For eigenstates written in the $SO(4)$-basis, $|\psi_{k}\rangle = \sum_{w=L}^{N} C_w^{(k)} |[N] w L\rangle_k$, we have $P^{(k)}_{SO(4)}$. Note that the sums in $n$ and $w$ are in increments of two units. 

When $\xi =0$, the eigenstates coincide with the $U(3)$-basis vectors and $P^{(k)}_{U(3)}=1$. As $\xi$ increases, the average level of delocalization in the $U(3)$-basis grows. However, the dependence of the values of $P^{(k)}_{U(3)}$ on the energies $E_k$ changes significantly for $\xi$ before and after the critical point. This can be seen in the five middle panels in Fig.~\ref{fig:DOS_IPR}, where we plot $P^{(k)}_{U(3)}/N$ \textit{vs} $E_k/N$. %When $\xi$ is still close to zero, there is a broad peak close to the middle of the spectrum. This peak moves to lower energies as the control parameter approaches $\xi_c$. 
Across the region $0<\xi \leq \xi_c$, $P^{(k)}_{U(3)}$ is a smooth function of energy. In contrast, a singularity appears above the critical point; a pronounced dip becomes noticeable at $E_{\text{ESQPT}}$. Its location moves as $\xi$ increases, following the increasing value of $E_{\text{ESQPT}}$. As seen for $\xi=0.6$, the dip becomes more pronounced as the value of $N$ increases.

The bottom panels of Fig.~\ref{fig:DOS_IPR} depict
$P^{(k)}_{SO(4)}/N$ \textit{vs} $E_k/N$ for different values
of $\xi$. The average level of delocalization in the $SO(4)$-basis
decreases as $\xi$ increases and $P^{(k)}_{SO(4)}=1$ when
$\xi=1$. At $\xi=0$, the plot shows a dip in the middle of the
spectrum with no relation to the ESQPT.  Below the critical point, as
$\xi$ increases, this dip moves toward lower energies and fades away. Above $\xi_c$, a discontinuity
appears at $E_{\text{ESQPT}}$. It is  much
less conspicuous than the dip in the $U(3)$-basis and it requires
large $N$ to be apparent, as seen in the inset for $\xi=0.6$. Notice also that $P^{(k)}_{SO(4)}$ peaks to its maximum value for the states with energies right above $E_{\text{ESQPT}}$. From these observations, one concludes that PR in the two bases can be used to
identify the ESQPT point, but it is by far more evident in the
$U(3)$-basis, especially for small $N$.

\begin{figure}[htb]
\centering
\includegraphics*[width=3.3in]{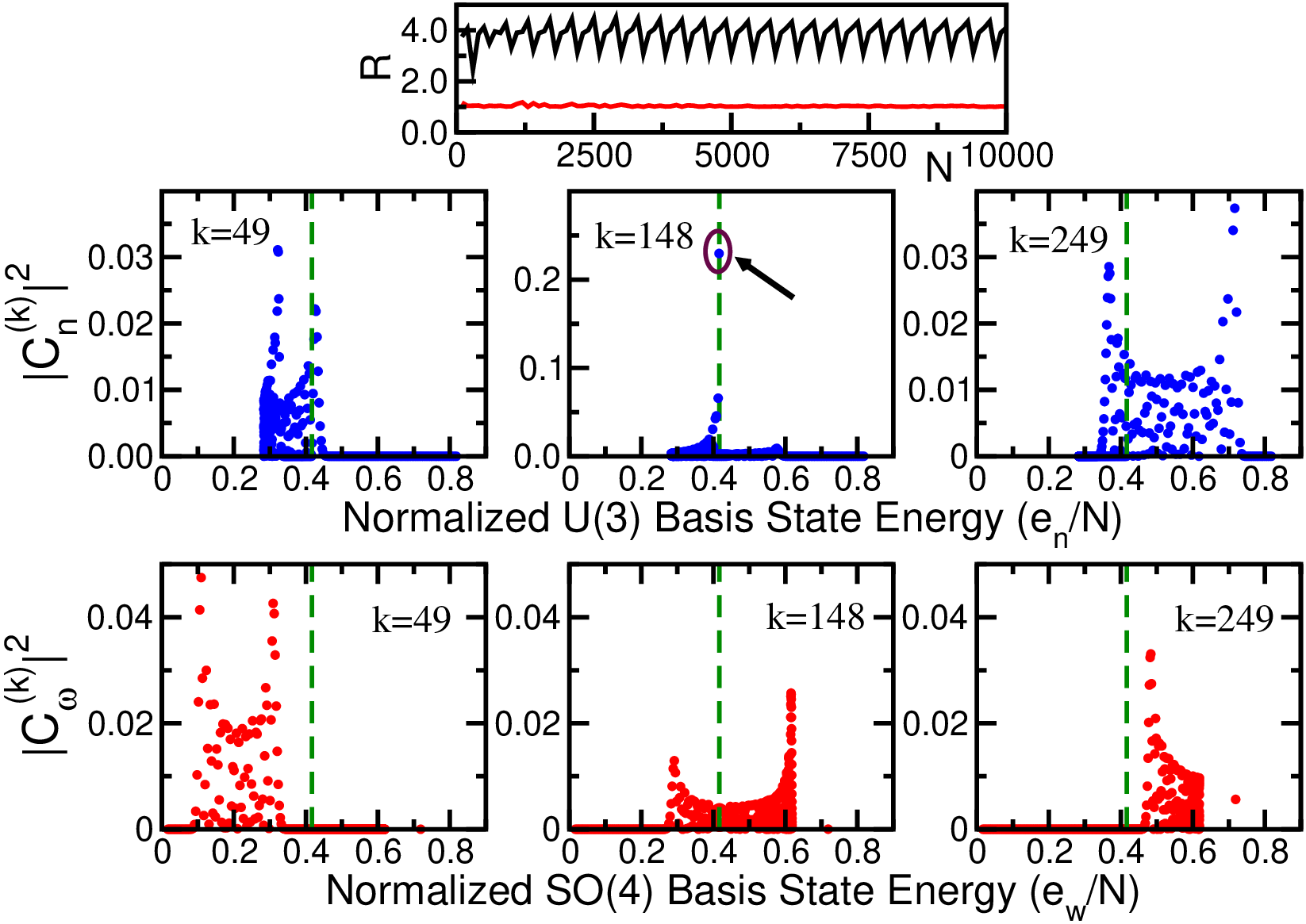}
\caption{(Color online) Top panel: ratio $R$ \textit{vs} $N$ for the eigenstate in the $U(3)$-basis closest to the separatrix (top curve) and for the one with the largest $P_{U(3)}$ (bottom curve). Middle and bottom panels: squared coefficients of the eigenstates,  respectively, in the $U(3)$- and $SO(4)$-basis \textit{vs} the energies of the corresponding basis vectors; $N=600$. All panels: $L=0$ and $\xi=0.6$. 
Left: low energy eigenstate $k=49$, $E_k/N=0.1767$. Middle: eigenstate with energy closest to the separatrix, $k=148$, $E_{\text{ESQPT}}/N=0.4173$. Right: high energy eigenstate $k=249$, $E_k/N=0.6513$. Vertical lines mark the ESQPT critical energy from Eq.~\eqref{eq:separatrix}. }
\label{fig:eigenstates}
\end{figure}

At first sight, the drop in the value of $P_{U(3)}$ and
$P_{SO(4)}$ at $E_{\text{ESQPT}}$ is counterintuitive. PR
usually reflects the density of states. Even though the values of PR
are intrinsically attached to a basis, in general, one expects the
states to be more extended where the density of states is larger. To
better understand what happens to the structure of the eigenstates in
the vicinity of $E_{\text{ESQPT}}$, we analyze in
Fig.~\ref{fig:eigenstates}, for $\xi=0.6$ and $N=600$, the
contributions to $|\psi_k\rangle $ from each basis vectors for three 
eigenstates, one before the ESQPT ($k=49$), the eigenstate closest to
the ESQPT ($k = 148$), and one after the ESQPT ($k =249$). We depict
the squared coefficients $|C_n^{(k)}|^2$ (middle panels) and
$|C_w^{(k)}|^2$ (bottom panels) as a function of the energies of the
bases, $e_b = \langle [N] b L | \hat{\cal H}_{U(4)} | [N] b L \rangle$ with $b = n$, $w$.
A similar study was done in Ref.~\cite{Caprio2008}, but the plots were for the coefficients with respect to the index of
the basis vectors, instead of their energies.

The structures of three selected eigenstates in the $U(3)$-basis
are illustrated in the middle panels of Fig.~\ref{fig:eigenstates}. For an
eigenstate with low energy ($k=49$), the largest contributions
(largest $|C_n^{(k)}|^2$) come from low values of $e_n$, with
a peak at each edge of the contributing energy interval. As $E_k$ increases and
approaches $E_{\text{ESQPT}}$, the largest value of $|C_n^{(k)}|^2$ on the
left of the contributing energy interval moves away from the boundary,
towards higher energies, and its amplitude increases substantially. At
the ESQPT ($k = 148$), this enhanced peak obscures the 
other components.  Compare the $y$-axis scale for $k = 148$ with $k=49, 249$. 
Finally, as $E_k$
further increases beyond the critical point, the enhanced peak
decreases and moves back to the left boundary of the contributing
energy interval, the whole interval naturally moving towards larger
values of $e_n$, as seen for $k=249$.

%Even though the eigenstates in the $U(3)$-basis with energy very close
%to the separatrix are extended in $e_n$, the preference for the first
%basis state, whose energy $e_{n=0}\sim E_{ESQPT}$, is bluntly
%evident. \textcolor{red}{Indeed, as seen in ???? of Fig.~\ref{fig:eigenstates}, 
The eigenstates in the $U(3)$-basis with energy very close to the separatrix have a blunt preference for the first basis state, whose energy $e_{n=0}$ is also $\sim E_{ESQPT}$. This is emphasized by the top panel of Fig.~\ref{fig:eigenstates}, which shows the ratio $R= |C_{n=0}/C_{n=\nu}|^2$ (where $|C_{n=\nu}|^2$ is the second largest component of the eigenstate) for various system sizes. $R$ is always larger than 1 for an eigenstate close to the separatrix (top curve). In contrast, $R \sim 1$ for eigenstates with large PR (bottom curve). The favoritism for $|[N] 0 ,0\rangle$ can be understood as follows. At the separatrix, the structure of the eigenstates
change from being close to the $SO(4)$ deformed symmetry to being more spherical [$U(3)$ DS].  The eigenstate $|\psi_{\text{ESQPT}} \rangle $ at the separatrix is the ground state of the spherical [$U(3)$] configuration, so its largest contribution comes from $e_{n=0}$. The high
level of localization of $|\psi_{\text{ESQPT}} \rangle $ in the
$U(3)$-basis can also be explained with dynamical considerations based
on the classical limit of the model~\cite{OurPreparation}.

The structures of the eigenstates in the $SO(4)$-basis are less
striking (bottom panels of Fig.~\ref{fig:eigenstates}). For an
eigenstate with low energy ($k=49$), the largest amplitudes of
$C_w^{(k)}$ occur at low values of $e_w$. As $E_k$ increases, the
contributing energy interval moves towards larger values of $e_w$.%, the
%largest $C_w^{(k)}$'s appearing at the boundaries of this interval.

%%%%%%%%%%%%%%%%%%%%%% DYNAMICS %%%%%%%%%%%%%%%%%%%%%%%

{\em Dynamics.--} We start by analyzing the effects of the ESQPT on the time evolution of initial states that correspond to $U(3)$-basis vectors, given the special role of this basis, as described above. Thus, $\hat{\cal H}_I=\hat{\cal H}_{U(3)}$ and $\hat{\cal H}_F=\hat{\cal H}_{U(4)}$. The simplest quantity to evaluate how fast an initial state $|\Psi(0)\rangle$ changes in time is the survival probability,
\begin{equation}
S_p(t) \equiv \left| \langle \Psi(0) | e^{-i \hat{\cal H}_F t} | \Psi(0) \rangle \right|^2  = \left| \sum_k |C_{k}^{(n)} |^2 e^{-i E_k t}  \right|^2,
\label{eq:fidelity}
\end{equation} 
where $C_{k}^{(n)}=\langle \psi_k | [N] n L \rangle$.  ${\cal P}(t)$ is the discrete Fourier transform in energy of the components $|C_{k}^{(n)} |^2$. Thus, the distribution of $E_k$ weighted by $|C_{k}^{(n)} |^2$ for a chosen initial state characterizes the decay of the survival probability~\cite{TorresAll2014}.

\begin{figure}[htb]
\centering
\includegraphics*[width=3.3in]{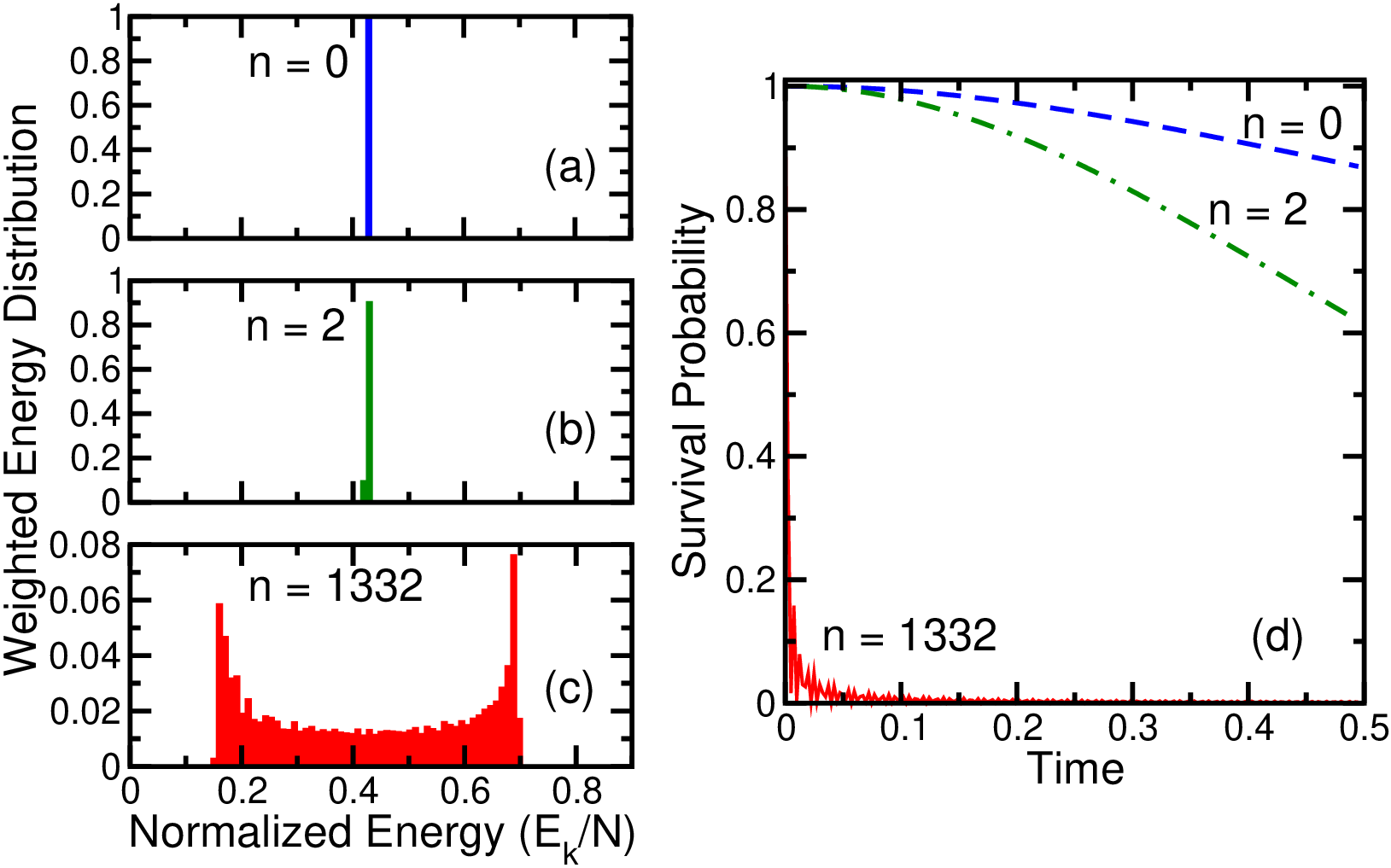}
\caption{(Color online) Weighted distribution of normalized excitation energies for three $U(3)$-basis vectors: $n = 0$ (a), $n = 2$ (b), and $n = 1332$ (c). In (d): Survival probability for the three basis vectors evolving according to  $\hat{\cal H}_{U(4)}$. $N=2000$, $L=0$, and $\xi=0.6$.}
\label{fig:LDOS}
\end{figure}

In Fig.~\ref{fig:LDOS} (a), we show the weighted energy distribution for an initial state with $n\!\!=\!\!0$.  Its energy, ${\cal E}_{n=0} = \langle [N] 0, 0 | \hat {\cal H}_F | [N] 0, 0 \rangle$, is the closest one to  $E_{\text{ESQPT}}$. As expected from the analysis of the structure of $|\psi_{\text{ESQPT}} \rangle$, the distribution is strongly localized at $E_{\text{ESQPT}}$. Thus, the evolution of this first $U(3)$-basis vector is very slow, as seen in Fig.~\ref{fig:LDOS} (d) and it does not accelerate as the system size increases~\cite{OurPreparation}. 

The second $U(3)$-basis vectors for $L=0$, $|[N] 2, 0 \rangle$, is slightly more spread out in energy [Fig.~\ref{fig:LDOS} (b)] than $|[N] 0, 0 \rangle$, its dynamics being then faster [Fig.~\ref{fig:LDOS} (d)]. As $n$ increases, the energy distribution stretches further in the direction of energies smaller and also larger than $E_{\text{ESQPT}}$. As a consequence, the second  state with ${\cal E}_{n}$ closest to $E_{\text{ESQPT}}$ is not $|[N] 2, 0\rangle$, but instead a high-$n$ basis vector with a more or less symmetric distribution around $E_{\text{ESQPT}}$ [Fig.~\ref{fig:LDOS} (c)]. Its decay is much faster than that of the first basis vectors [Fig.~\ref{fig:LDOS} (d)] and, for the system sizes studied, it gets faster as $N$ increases~\cite{OurPreparation}. %At very high values of ${\cal E}_{n}$, the distribution localizes again. For $n\sim N$, the largest contribution comes from the right edge of the spectrum and the dynamics is once more very slow, in fact even slower than for $n\sim0$ (not shown).

Similarly to what happens to the structure of the eigenstates in the $SO(4)$-basis, the projection of one $|[N] w L\rangle$ onto
$|\psi_k\rangle$ has an `accordion-like' behavior as ${\cal E}_w $ increases from zero. The energy distribution of $E_k$ weighted by $|C_w^{(k)}|^2$ is localized around low (high) values of $E_k$ when ${\cal E}_w$ is small (large) and it spreads out for ${\cal E}_w$ away from the edges of the spectrum. The dynamics reflects these distributions, being slower and the fluctuations after saturation being larger for initial states with energies closer to the border of the spectrum than for those away from it. No special behavior is observed for ${\cal E}_w \sim E_{\text{ESQPT}}$.

%%%%%%%%%%%%%%%%%%% CONCLUSIONS %%%%%%%%%%%%%%%%%%%%%%%%%%

{\em Experimental realization.--} Recent experiments with trapped ions~\cite{Jurcevi2014,Richerme2014} studied the quench dynamics of systems where the range of the interaction was tunable. One of the systems considered was described by the Ising spin-1/2 Hamiltonian with a transverse field, which in the limit of infinite-range interaction corresponds to the LMG model~\cite{Cejnar2009,Fernandez2009,Yuan2012}. Thus, the experimental setup to compare the speed of the evolution for different basis vectors is already available.

{\em Conclusion.--} In general, the dynamics of initial states with energies very close to the edges of the spectrum is much slower than
for states with energies closer to the middle of the spectrum~\cite{TorresAll2014}. Here, we showed that in an isolated system undergoing an ESQPT, a slow time evolution can occur also for initial states with energy very close to $E_{\text{ESQPT}}$. This is
the case of an initial state corresponding to the ground state of the $U(n)$ subalgebra of a model described with a $U(n+1)$ algebraic structure. This behavior reflects the structures of the eigenstates of the $U(n+1)$ Hamiltonian close to the separatrix, which are highly localized in the ground state of the $U(n)$ Hamiltonian. Our
findings have therefore identified two additional methods to detect the presence of an ESQPT, by analyzing the level of delocalization of
the eigenstates in the $U(n)$-basis and by comparing the speed of the evolution of different $U(n)$-basis vectors evolving under $U(n+1)$ Hamiltonians, the latter being more accessible experimentally.

We plan to extend the present studies to consider also the effects of an ESQPT associated with first order phase transitions~ \cite{macek2014,stransky2014}. We also intend to study the influences of ESQPTs on coupled systems \cite{stransky2014}.

%%%%%%%%%%%%%%%%%%%% ACKNOWLEDGMENTS %%%%%%%%%%%%%%%%%%%%%

\begin{acknowledgments}
LFS was supported by the NSF grant No.~DMR-1147430. FPB was funded by MINECO grants FIS2011-28738-C02-02 and FIS2014-53448-C2-2-P and by Spanish Consolider-Ingenio 2010 (CPANCSD2007-00042). LFS thanks the ITAMP hospitality, where part of this work was done. We thank Jos\'e M. Arias, Jos\'e E.\ Garc\'ia-Ramos, Francesco Iachello, and Pedro P\'erez-Fern\'andez for useful discussions.
\end{acknowledgments}

%%%%%%%%%%%%%%%%%%%% REFERENCES %%%%%%%%%%%%%%%%%%%%%

\end{document}